\documentclass[12pt]{article}
\usepackage{latexsym}
\usepackage{amssymb}
\usepackage{amsfonts}

\usepackage{color,graphicx} 
\definecolor{brownn}{cmyk}{0,1,1,0.5}
\definecolor{bluen}{rgb}{.1 ,0, .8}

\usepackage{ifpdf}
\ifpdf
\RequirePackage[T1]{fontenc}
\RequirePackage{times}
\usepackage[hypertexnames=false,
colorlinks,linkcolor=bluen,citecolor=brownn,urlcolor=brownn]{hyperref}
\usepackage{epstopdf} 
\else
\usepackage[hypertex,hypertexnames=false,
colorlinks,linkcolor=bluen,citecolor=brownn,urlcolor=brownn]{hyperref}
\fi
\graphicspath{{images/}}

\def\nc#1{\newcommand{#1}}
\def\rnc#1{\renewcommand{#1}}

\def\b{\beta}
\nc{\g}{\gamma}

\nc{\D}{\Delta} 
\nc{\e}{\eta}
\nc{\ep}{\epsilon}

\nc{\ve}{\varepsilon}
\nc{\G}{\Gamma}

\nc{\la}{\lambda}
\nc{\La}{\Lambda}
\nc{\om}{\omega}
\nc{\Om}{\Omega}
\nc{\vphi}{\varphi}
\nc{\si}{\sigma}
\nc{\Si}{\Sigma}
\rnc\th{\theta}
\nc\Th{\Theta}
\nc{\z}{\zeta}

\nc{\got}[1]{\mathfrak{#1}} 

\def\det{{\rm det}}

\nc\im{{\rm Im}\, }
\nc\re{{\rm Re}\, }

\nc{\Rt}{{\tilde R}}
\nc{\CC}{{\mathbb C}}
\nc\II{{\mathbb I}} 
\nc{\RR}{{\mathbb R}}
\nc{\HH}{{\mathbb H}}
\nc{\NN}{{\mathbb N}}
\nc{\ZZ}{{\mathbb Z}}
\nc{\MM}{{\mathbb M}}

\nc{\ov}[1]{\overline{#1}}

\nc{\non}{\nonumber\\}

\nc{\noi}{\noindent}

\nc{\p}{\partial}
\nc{\na}{\nabla}
\def\x{\times}

\nc\vev[1]{\ensuremath{\lan #1\ran} {}}
\nc\refeq[1]{(\ref{#1})}
\nc{\eqref}[1]{(\ref{#1})}
\rnc\to[1][]{\ensuremath{\stackrel{#1}{\rightarrow\;}}}

\nc{\twovec}[2]{\left( \!\!
\begin{array}{c} #1\\  #2 \end{array}\!\!\right)}
\nc{\twomat}[4]{\left(\!\! \begin{array}{cc} #1&#2\\ 
#3&#4\end{array}\!\! \right)}

\nc{\ds}{\displaystyle}

\nc{\lan}{\langle}
\nc{\ran}{\rangle}

\nc{\beq}{\begin{equation}}
\nc{\eeq}{\end{equation}}
\nc{\beqa}{\begin{eqnarray}}
\nc{\eeqa}{\end{eqnarray}}
\nc{\beqas}{\begin{eqnarray*}}
\nc{\eeqas}{\end{eqnarray*}}
\nc{\barr}{\begin{array}}
\nc{\earr}{\end{array}}
\nc{\ben}{\begin{enumerate}}
\nc{\een}{\end{enumerate}}
\nc{\bit}{\begin{itemize}}
\nc{\eit}{\end{itemize}}

\nc{\cred}{\color{red}}
\nc{\cblue}{\color{blue}}

\nc\rQ[1][]{\ensuremath{{\cred\leftarrow(?)\mbox{\footnotesize #1}} } }

\nc\more{{ \cred{MORE}}}
\nc{\remark}[1]{\cblue[Rem]\footnote[*]{\color{blue}{Remark:} #1}}
\nc{\foot}[1]{{}{\cblue[{}\footnote[0]{\tt #1}]}}
\nc{\ADD}{{\cblue ADD}}

\nc\fb{\ensuremath{{\bar 5}} {}}
\nc\yfb{\ensuremath{Y_{\overline 5}} {}}
\nc\yf{\ensuremath{Y_5} {}}
\nc\yt{\ensuremath{Y_{10}} {}}
\nc\ytb{\ensuremath{Y_{\overline{10}}} {}}
\nc\y[2][]{\ensuremath{Y_{#2}^{#1}} {}}
\nc\f[2][]{\ensuremath{5_{(#2)}^{#1}} {}}
\nc\hfb[1][]{\ensuremath{{\bar 5}^{\, #1}_H} }
\nc\yu[2]{\ensuremath{y_{#2}^{(#1)}}{}}

\definecolor{lightgray}{cmyk}{0.1,0.2,0,0.1}

\nc\lgut{\La_{GUT}}
\nc\mpl{M_{Pl}}
\nc\fdx{r_{D/X}}

\normalsize

\begin{document}
\thispagestyle{empty}
\title{A  model of Yukawa couplings with matter-messenger
 unification} 
\author{ J. Pawe{\l}czyk\\{\small {}}\\{\small {\it Institute of 
Theoretical Physics, Faculty of Physics, University of Warsaw,}}\\{\small {\it 
Ho\.za 69, 00-681 Warsaw, Poland}}\\}
\date{}
\maketitle
\abstract{\normalsize
We propose a GUT model in which  visible  matter  and  messengers    are treated in unified  way	what unavoidably leads to messenger-matter mixed   Yukawa interactions. 
Influence of this mixing on the  fermion masses and  the weak mixing angles is discussed. 

\newpage
\normalsize

GUT models are attractive scenarios for unification of all gauge forces
\cite{Langacker:1980}. 
The basic arguments supporting the idea are twofold: all known 
fermionic matter is organized in SU(5)  multiplets and coupling constants seem to unify at some scale.
It  appears that supersymmetric versions of these models  provide  better coupling unification and shift the unification scale  $\La_{GUT}$ to values acceptable for the proton stability  under heavy gauge bosons exchange \cite{Kounnas:1983,Nanopoulos:1983}. 
Besides many nice features GUT  models face serious issues among which there is the flavour problem i.e. the problem of generating appropriate hierarchy between masses of fermions. For example
the simplest GUT models require $m_i^d=m_i^l$ at the GUT scale for each family ($i=1,2,3$) of down quarks ($d$) and leptons ($l$). 
Calculations shows that one can hope to match masses of fermions  of the 3rd family, $m_\tau=m_b$ \cite{mb-mtau,elor}, but this can not be achieved  for the other two families. The situation is readily apparent when one compares one-loop RG independent ratios of the light lepton and quark masses: $m_e/m_\mu\approx 0.004,\ m_d/m_s\approx 0.05$ what means that $m_e/m_d\sim 9\, m_\mu/m_s$. 
It is clear that an extra source of  flavour symmetry breaking is needed. The subject has very long history \cite{Georgi:1979df} and was studied very extensively in the past, see e.g. \cite{fl-GUT} and 
\cite{pokorski} and references contained therein.

The motivation behind our approach to the flavour problem lays in some F-theory constructions of GUT theories (the so-called F-GUTs)
\cite{F-unify} where the  both flavour visible chiral matter and 
messengers  originate from the same 
D7-brane intersection \cite{HV-E8} (see the Appendix).
This means that matter and 
messengers\footnote{We shall abuse the name "messenger" here because the SUSY breaking will not discussed here.}
 should be treated on equal  footing e.g.  should have common Yukawa matrix. Mixed couplings between fields is generated by fluxes on Calabi-Yau manifold \cite{Yukawa-HV,Yukawa-np}.
We shall claim  that  the mixing is responsible for the deviation from the naive relations between fermion masses.

We begin writing down all  relevant superpotential terms 
\beq\label{mainY}
(y^u)_{ij}10_i10_j (5_H)_2+(\tilde y^d)_{iJ}10_i\fb_J (\fb_H)_2+
a_I \fb_I 5 X
\eeq
 where the subscript 2 attached to the Higgs fields recalls that we take into account only the doublet   part of  the 5's.
The above superpotential  in principle could be extended with extra $10$'s but it  is unlike in F-theory setup (see the Summery).
Thus we let the flavour indices run as follows: $i,j=1,2,3;\; I,J=1,...4$ and we add an extra $5$ which will form the messenger vector pair with one of the \fb's.
The spurion v.e.v. $\vev X$  gives mass $M_Y$ to the messengers (and could trigger SUSY breaking through its nontrivial F-term).
It is clear that \vev X provide mass $M_Y$ only to one vector pair $(5,\yfb)$ where $\yfb\sim a_J \fb_J $ leaving the other \fb matter massless. 
Below the scale $M_Y$ we are left with 3 massless families. Accordingly we need to redefine fields what results in modification of Yukawas $\tilde y_d$ \footnote{For other GUT models utilizing mixing of matter with some extra fields see \cite{babu-y-mix} and references therein.}.

F-theory GUTs \cite{Yukawa-HV}\footnote{For different approaches to fermion hierarchies in F-theory GUTs see \cite{Yukawa-np, Yukawa-others}.}
 lead to  Yukawa matrices  of the Froggatt-Nielsen form \cite{Froggatt-Nielsen}
possessing naturally hierarchical structure. 
We write explicitly the general structure of these couplings in the case of three 10's and four \fb but the matrices can be easily made up for the other cases.
\beq\label{ync}
y_{NC}\sim\left(
\barr{cccc}
h_\th^5&h_\th^4& h_\th^3& h_\th^2\\
h_\th^4& h_\th^3& h_\th^2& h_\th\\
h_\th^3& h_\th^2& h_\th& 1\\
\earr
\right)+
\Phi_0 \left(
\barr{cccc}
h_\th^4&h_\th^3& h_\th^2& h_\th\\
h_\th^3& h_\th^2& h_\th& 0\\
h_\th^2& h_\th& 0& 0\\
\earr
\right)+
\Phi_0^2 \left(
\barr{cccc}
h_\th^3&h_\th^2& 0& 0\\
h_\th^2& 0& 0& 0\\
0& 0& 0& 0\\
\earr
\right)
\eeq
where $h_\th, \Phi_0$ are free parameters and as usual the order one constants have been dismissed.
It is known  that $\Phi_0$ depends on the hypercharges of the corresponding matter. For relatively big hypercharges one expects
$\Phi_0=0$.  Thus this case should  be  applicable to leptons and to some extend to up-quarks\footnote{Recall values of the hypercharge for R component of fermions: $Y(u_R)=2/3,\ Y(e_R)=-1,\ Y(d_R)=-1/3$.}. 
We want to stress that $\tilde y^d$ dependence on the hypercharge
of matter breaks SU(5) symmetry so in fact Yukawa coupling are different for down quarks and leptons. 

The model with just 3 families of matter and no matter-messenger mixing (thus disregarding advocated here messenger i.e. the last column of \refeq{ync}) was discussed in \cite{Yukawa-HV}. For $h_\th\sim \ep^2,\ \ep\sim 0.2$ this  model 
is fine for the up quarks ($m_u:m_c:m_t\sim \ep^8:\ep^4:1$)
 but fails for seemingly more favorable case of leptons due to 
$m_e/m_\mu\approx 0.005\sim \ep^{3.5},\ m_\mu/m_\tau\approx 0.05\sim\ep^2$. Recall that in MSSM ratios of Yukawas for fermions differing only by the family index depend on scale mildly  thus  RG flow can not explain this deviation. 
For down quarks one puts instead $\Phi_0\sim \ep\sim 0.2$ \cite{Yukawa-HV}. This  yields some misfit for masses $m_1:m_2:m_3=\ep^6:\ep^3:1$ and  correctly reproduces 
mixing angles $\th_{23}\sim \ep^2,\ \th_{13}\sim\ep^3$
although $\th_{12}\sim \ep^2$ is too small.

Now we discuss how the situation  changes if we go the model 
\refeq{mainY} with an extra \fb. Its important to notice that 
the last term of \refeq{mainY} change the definition of messenger by mixing  
\fb's.
One expects to get some kind of hierarchy between $a_I$s e.g. of the \refeq{ync} type (take just the last row) i.e. one expects
that $a_{I+1}\ll a_I$. Here we normalize  $a_4=1$.  The necessity to redefine messenger leads to redefinition of the other fields in multiplets of \fb. This rotates Yukawa coupling matrix. On the other hand the leading terms of the couplings between the physical messenger \yfb and light matter  has universal form. 
\beq
(y_{14} \ep^4 10_1+y_{24} \ep^2 10_2+ y_{34} 10_3)\cdot \yfb\cdot (\fb_H)_2
\eeq
After decoupling of the messenger \yfb
we are left with 3$\x$3 matrix of Yukawas for visible fermions. Below we analyze this matrix for the cases appropriate for up and down quarks and for leptons.

\paragraph{Leptons. } 
Putting all the unknown constants in \refeq{ync} in the leading order in $\ep$ we get for leptons ($\Phi_0=0$ case):
\beq\label{yl34}
\tilde y^l = \left(
\barr{cccc}
y_{11} \,\ep^{10}& y_{12} \,\ep^8& y_{13} \,\ep^6& y_{14} \,\ep^4\\
 y_{21} \,\ep^8& y_{22} \,\ep^6&    y_{23} \,\ep^4& y_{24} \,\ep^2\\ 
   y_{31} \,\ep^6& y_{32} \,\ep^4& y_{33} \,\ep^2& y_{34}
\earr
   \right)
\eeq
Properties of the resulting 3$\times$3  matrix (called here $y^l$) for massless fermions  depends on  $a_I$ coefficients.
For the sake of illustration of possibilities at hand we display formulae for $a_1\gg \ep^6$ and $a_2\gg a_3 \ep^2$.
 The resulting Yukawa  couplings for leptons  is
\beq\label{yl-e}
y^l= 
\left(
\barr{ccc}
y_{14}a_1^*\,\ep^4 &  y_{14 }a_2^*\,\ep^4 + y_{12}\,\ep^8  &  y_{14}a_3^*\,\ep^4 +  y_{13}\,\ep^6 \\
  y_{24}a_1^*\,\ep^2 & y_{24 }a_2^*\,\ep^2  +  y_{22} \,\ep^6&
  y_{24}a_3^* \,\ep^2  +  y_{23}\,\ep^4\\ 
a_1^* y_{34} &a_2^* y_{34} + y_{32}\,\ep^4 & a_3^* y_{34} +  y_{33}\,\ep^2
\earr\right)
\eeq
Diagonalization of \refeq{yl-e} yields  the following
eigenvalues (given here up to phases):
\beqa\label{yl-diag}
y_3&\approx &a_3^* y_{34} +  y_{33}\,\ep^2\\
y_2&\approx &\frac{ (y_{24} y_{33} - y_{23} y_{34}) a_2 \ep^4}{a_3^*y_{34} +
  y_{33}\,\ep^2}\\
y_1&\approx &\frac{\det{(y_d)}}{ (y_{24} y_{33} - y_{23} y_{34}) a_2 \ep^4}
\eeqa
where
$
\det{y^{\,l}}\sim a_1^*\,\ep^{12} 
$.
Assuming generic values of $y_{ij}\sim 1$ one gets
\beq
\frac{y_1}{y_2}\sim \frac{a_1 y_3 \ep^4}{a_2^2},\quad \frac{y_2}{y_3}\sim 
\frac{a_2\ep^4}{y_3^2}
\eeq
Needed  lepton hierarchy $y_1/y_2\sim \ep^{3.5},\ y_2/y_3\sim \ep^2$ yields $a_2 \ep^2\sim y_3^2$ and $a_1 y_3 \sim a_2^2\ep^{0.5}$.
 For $y_3\sim \ep^k$ this gives $a_2\sim\ep^{2k-2}$ and $a_1\sim \ep^{3k-4.5}$.
 Thus hierarchy in $a_I$'s is respected if $k\geq 2$, e.g.   $y_3\sim \ep^{2.5}$  produces $a_2\sim \ep^3,\ a_1\sim\ep^{3}$.
 This choice of $y_3$ requires some fine tuning i.e. $y_{33}\sim \ep^{0.5}\sim 0.5$ and $a_3\sim \ep^{2.5}$.

\paragraph{Down quarks.} Now we turn to the second example $h_\th\sim\ep^2$ and $\Phi_0\sim\ep$ which we use to fit to down quark Yukawas.
Repeating the previous construction 
instead of \refeq{yl-e} one obtains\footnote{In general the coefficients $y_{ij}$ in matrices $y^l$ and $y^d$ can be different.}:
\beq\label{yd-e}
y^d=\left(
\begin{array}{ccc}
 y_{11} \ep^8+y_{14} a_1^* \ep^3 & 
y_{12} \ep^6+y_{14} a_2^* \ep^3 & y_{13} \ep^5+y_{14}a_3^* \ep^3 \\
 y_{21} \ep^6+y_{24} a_1^* \ep^2 & 
y_{22}\ep^5+y_{24} a_2^* \ep^2 & y_{23} \ep^3+y_{24}a_3^* \ep^2 \\
 y_{31} \ep^5+y_{34} a_1^* & 
y_{32}\ep^3+y_{34} a_2^* & y_{33} \ep^2+y_{34}a_3^* \\
\end{array}
\right)
\eeq
The obtained diagonalized Yukawas are (for $a_1\gg \ep^5,\ a_1\geq\ep^2 a_2,\ a_1\geq\ep^3 a_3$)

\beqa\label{y-diag}
y_3&\approx &a_3^* y_{34} +  y_{33}\,\ep^2\\
y_2&\approx &\frac{  a_2 y_{23} y_{34} \ep^3 +  y_{32} y_{33} \ep^6}{y_3}\\
y_1&\approx &\frac{\det{(y_d)}}{ a_2 y_{23} y_{34} \ep^3 +  y_{32} y_{33} \ep^6}
\eeqa
where
$\det(y_d)\approx(y_{12}y_{23}y_{34}-y_{14}y_{23}y_{32})a_1\ep^9$.
Phenomenological values of the ratios of the down quark masses
 $m_d:m_s:m_b=\ep^{2+s}:\ep^s:1$ for $s=2-3$.
With $y_3\sim \ep^k,\ a_2\sim \ep^{2k+s-3},\ a_1\sim \ep^{3k+2s-7}$ we get correct relations for down quarks Yukawas. 
The coefficients $a_I$ have hierarchical structure if $k+s> 4$.
One can also calculate the mixing angles: $
\th_{12}\sim \ep^{6-(k+s)}+\ep^{k},\ \th_{23}\sim \ep^{3-k}+
\ep^{2},\ \th_{13}\sim \ep^{5-k}+\ep^{3}$ (coefficients of the order one are suppressed).
 e.g. for $k=1.5,\ s=3$ one has : $\th_{12}\sim \ep^{1.5},\ \th_{23}\sim \ep^{1.5},\ \th_{13}\sim\ep^3$. These values match phenomenology up to terms of the order $O(\ep^{0.5})$. 

\paragraph{RG flow of masses.} 
Finally we emphasize that 
the presented results hold at the GUT scale and are subject to
 renormalization what in case of matter-messenger coupling may
 have important influence on ordinary Yukawas.
 In order to analyze the problem we did numerical calculation based on one-loop RG equations for the wave functions  renormalizations  presented in \cite{Chacko-Ponton} and corrected for the case at hand in \cite{Shih-soft-m}. It RG results depends mainly on the messenger mass $M$ and the largest mixing coupling $y_{34}$.
The RG flow modification of mixing angles is negligible.
\begin{table}[h]
$$
\barr{|c||ccc|ccc|ccc|}
\hline
&	m_u & m_c & m_t & m_d & m_s & m_b & m_e & m_{\mu } &m_{\tau }\\[-.2cm]
&	{}_{[MeV]} & {}_{[MeV]} & {}_{[GeV]} & {}_{[MeV]} & {}_{[MeV]} & {}_{[GeV]} & {}_{[MeV]} & {}_{[MeV]} &{}_{[GeV]}\\
	\hline
MSSM, \La=10^3&	1.15 & 560 & 161 & 2.20 & 42.0& 2.23 & 0.41 & 88.3 & 1.50\\
	\hline\hline
MSSM&	0.55 & 270 & 106 & 0.79 & 15.1 & 0.89 & 0.29 & 62.0 & 1.06\\
	\hline
y_{34}=0.6, M=10^{13}&	0.55 & 268 & 107 & 0.84 & 16.0 & 0.96 & 0.31 & 66.2 & 1.16\\
	\hline
	y_{34}=1.5, M=10^7&	0.51&250&112& 1.39&26.6&1.82& 0.52&112&2.48\\
	\hline
\earr
$$\centering
\parbox{.8\textwidth}{\caption{One-loop fermion masses at $\La_{GUT}$ for different values of the messenger-matter mixing $y_{34}$ and the messenger 
mass $M$. The initial values of the masses at 1 TeV  are given in the first row.}}
\end{table}
The results are presented in Table 1. It is clear that  RG does not change the fermion masses ratios in a significant way.

\paragraph{Summary.} Yukawa couplings for quarks and leptons can be modified by mixing with extra vector matter called messengers here. This mechanism opens up an extra path toward obtaining the desired structure of the  Yukawa couplings of the visible matter. In the present paper we have worked out an example based on a model inspired by F-theory for which the possible Yukawa  couplings have  form of the Froggatt-Nielsen type. 
The considered vector pair of matter was $5+\fb$ of SU(5). 
Extending the model by an extra $10$'s is excluded
if one wants to stay within perturbative physics. The reason is that due to  
hierarchical structure of the proposed Yukawa couplings and the couplings $a_I$'s to spurion \refeq{mainY} the extra $10$'s would have couplings hierarchically much bigger then that of the top quark i.e. much bigger then 1. 
Hence the messenger sector would be strongly interacting. 
This might be appealing in the F-theory context but usually is not welcomed feature.  

From \refeq{y-diag} one has $y_b\sim a_3^* y_{34} +  y_{33}\ll y_{34}\sim 1$ what implies that $\tan\b$ can not be large (typically smaller than 10) in this type of models. Adding more vector pairs of $5$'s  would lower $\tan\b$ much further what is excluded in MSSM--like models \cite{Draper}.
Thus we infer that 
our scenario prefers the single messenger in the representation $5$ of SU(5) GUT.
It is also interesting that \cite{HV-E8} do not contain any model with Majorana neutrinos masses which would fit in our scheme but numerous with Dirac masses. This is shortly discussed in the appendix.

We want to emphasize here that the results obtained shows only the order of quantities of interest in terms of powers of $\ep$. But because $\ep\sim 0.2$ and unknown coefficients $y_{ij}\sim 1$ this expansion may be misleading e.g. if one takes some $y_{ij}\sim 0.5$ than  terms quadratic in these  $y$'s are of the order $\ep$ what spoils the expansion we have made.


The  presented model has various phenomenological consequences. The subject is very reach and requires separate studies so we shall be very brief here.
Strong ($y_{34}^d\sim 1$) mixing of messengers 
to top quark 
produces  large $A_t$ term \cite{G-R}. 
\beq
V=-\frac{F_X}{16\pi^2\vev X}
(y_{34}^d)^2\,y_t\, \tilde Q H_u \tilde t
+c.c.
\eeq
$A_t$  of the order of several TeV  gives positive contribution to  the  Higgs mass \cite{Martin-rev,JPT} helping to achieve required value 126 GeV. 
Moreover it decreases the lighter stop mass  \cite{Drees-book}. Both contributions are very welcomed in view of LHC results: 
stop can still be relatively light (few hundred GeV) thus its discovery would  favor large $A_t$ term.  

The other A-terms are much smaller than $A_t$ but nevertheless are severely constrained. The reason is that 
flavour-messenger mixing yields non-diagonal soft masses \cite{Chacko-Ponton,Shih-soft-m}
which in turn generate FCNC. Phenomenological  constrains on FCNC \cite{Masiero-FCNC-rev} 
put limits on mixing coefficients $y_{i4}$. 
The thorough  analysis of this problem is left to our future work \cite{jpt-mixing}.

\appendix
\section{About F-theory constructions}\label{app}
Here we discuss SU(5) F-theory local models of \cite{HV-E8} which lead to the scenario proposed in the current work. In F-theory various matter is localized on different Riemann surfaces along which D7-banes intersect.
Fluxes on these Riemann surfaces determines number of chiral species of given matter. For all models we need the flux to prevent existence of \ytb
 so all $10$'s are flavours.
We also need appropriate $\yf$ and the spurion $X$.
In what follows we shall assume that the appropriate fluxes can be switched on.

The group theory properties of the matter follows from $SU(5)_\perp/\G$ (which is the global symmetry group here) for various discrete groups $\G$. In \cite{HV-E8} scenarios for different $\G$'s has been considered.
Here we shall shot-cut the discussion to the relevant pieces of the models.
We start with Dirac scenario models. For $\G=\ZZ_2$ the global symmetry group is $U(1)_{PQ}\times U(1)_{\chi}$. All  possible versions these models realizing scenarios proposed in the present work are enumerated in Table 2.

\begin{table}[h]
\centering
\begin{tabular}{|c|c||c|c||c|c||c|c|}
\hline
&&\multicolumn{2}{c||}{version 1}&\multicolumn{2}{c||}{version 2}&\multicolumn{2}{c|}{version 3}\\
\hline
&\fb & \yf=\f1 & $X=N_R$& \yf=\f2 & $X=D_{(2)}$& \yf=\f3 & $X=D_{(4)}$ \\\hline
$U(1)_{PQ}$ & $+1$  & $+2$ & $-3$& $-5$ & $+4$& $+6$ & $-7$\\\hline
$U(1)_{\chi}$ & $+3$ & $+2$ & $-5$& $-3$ & $0$& $+2$ & $-5$\\\hline
\end{tabular}
\hspace{15mm}\parbox{15.5cm}{\caption{$U(1)_{PQ}$ and $U(1)_{\chi}$
charges for the candidates for messengers $Y_5$ and spurions $X$ of Eq.\refeq{mainY} for different variants of $SU(5)_\perp/\ZZ_2$ model. The \fb has universal charges.
}}
\end{table}
For $\G=\ZZ_2\x\ZZ_2$ only the last model of Table 2 survives.
For $\G=\ZZ_3$ there is one possibility for which all the relevant fields are the same as  originally listed in \cite{HV-E8}.
On the other hand non of  Majorana models of \cite{HV-E8} can realize the scenario of the present paper.

\section*{Acknowledgments}
\vspace*{.5cm}
\noindent The author would like to acknowledge fruitful remarks of Krzysztof Turzy\'nski on the manuscript and
stimulating discussions with 
Tomasz Jeli\'nski,  Stefan Pokorski and Krzysztof Turzy\'nski.

\noindent This work was partially supported by the 
Polish Ministry of Science MNiSW grant
under contract N N202 091839 (2010-2013).

\end{document}